\documentclass[12pt,preprint]{aastex}
\usepackage{epsfig} 
\usepackage{natbib}
\newcommand{\etal}{\mbox{\rm et al.~}}
\newcommand{\ms}{\mbox{m s$^{-1}$}}

\received{}
\revised{}
\accepted{}

\slugcomment{submitted to ApJS}

\shorttitle{The Lick Observatory Planet Search}
\shortauthors{Fischer \etal}

\begin{document}

\title{The Twenty-Five Year Lick Planet Search$^{1}$}

\author{Debra A. Fischer\altaffilmark{2},
	Geoffrey W. Marcy\altaffilmark{3,4},
	Julien F. P. Spronck\altaffilmark{2} }

\email{debra.fischer@yale.edu}

\altaffiltext{1}{Based on observations obtained at the Lick Observatory, which is 
operated by the University of California} 

\altaffiltext{2}{Department of Astronomy, Yale University, New Haven, CT USA 06520}

\altaffiltext{3}{Department of Astronomy, University of California,
Berkeley, CA USA  94720-3411}
  
\altaffiltext{4}{Department of Physics and Astronomy, San Francisco State University,
San Francisco, CA, USA 94132}

\altaffiltext{5}{UCO/Lick Observatory, 
University of California at Santa Cruz, Santa Cruz, CA, USA 95064}

\begin{abstract}
The Lick planet search program began in 1987 when the first spectrum of $\tau$ Ceti 
was taken with an iodine cell and the Hamilton Spectrograph. 
Upgrades to the instrument improved the Doppler precision from about 10 \ms\  in 1992 to about 
3 \ms\ in 1995. The project detected dozens of exoplanets with orbital periods ranging from a few days to 
several years. The Lick survey identified the first planet 
in an eccentric orbit (70 Virginis) and the first multi-planet system around a 
normal main sequence star (Upsilon Andromedae). These discoveries advanced our understanding 
of planet formation and orbital migration. Data from this project helped to quantify 
a correlation between host star metallicity and the occurrence rate of gas giant planets. 
The program also served as a test bed for innovation with testing of a tip-tilt system
at the coud{\'e} focus and fiber scrambler designs to stabilize illumination of the spectrometer
optics.  The Lick planet search with the 
Hamilton spectrograph effectively ended when a heater malfunction compromised 
the integrity of the iodine cell. Here, we present more than 14,000 velocities for 
386 stars that were surveyed between 1987 and 2011. 
\end{abstract}


\section{Introduction}
\label{intro}
Humans have long been curious about the existence of planets orbiting other 
stars and steady technological advances over the past half century resulted in  
remarkable progress that enabled the detection of hundreds of exoplanets.  
\citet{S52} was first to suggest that even then state of the art radial velocity measurements 
might be used to detect putative gas giant planets if they orbit close to their host stars. 
Twenty years later, \citet{GG73} outlined a strategy for achieving high Doppler precisions; they 
emphasized that the key was to superimpose a wavelength standard in the stellar spectrum 
so that instrumental shifts would equally affect both the calibrator and the stellar observations. 
They pointed out that telluric lines could serve in this role and estimated that Doppler shifts 
of 10 \ms could be measured with the limiting precision set by 
pressure, temperature and wind variations in the Earths atmosphere. 

\citet{CW79} designed an ingenious version of telluric lines in a bottle, filling
a 0.90-m long glass cell with hydrogen fluoride (HF) gas and positioning the cell in front 
of the slit of a coud{\'e} spectrograph at the 3.6-m Canada France Hawai'i telescope 
(CFHT).  This pioneering technique was ideal for a general purpose spectrograph 
because the HF absorption lines were superposed on the stellar spectrum. By virtue 
of tracing out an identical light path, the HF wavelength calibration was imprinted with 
the same instrumental profile as the stellar spectrum. The team launched a survey of 
16 stars during 27 observing runs (of three or fewer nights each) spanning six years. 
They used a spectrograph flux meter to derive accurate photon-weighted midpoints 
for their exposure times and calculated run-to-run corrections of tens of meters 
per second for zero-point variations \citep{CWY88}. The team also modeled the effect of 
asymmetry variations in the instrumental profile of the spectrograph 
\citep{CW85}. In this sample of 16 stars, \citet{CWY88} identified 
residual velocity variations to the binary orbit 
of $\gamma$ Cephei \citep{W92}. Although this velocity scatter 
was initially attributed to stellar noise sources, \citet{H03} later determined that these 
were reflex velocities induced by a gas giant planet. Fourteen stars 
showed internal errors of about 10 \ms\ and rms scatter of about 13 \ms.  
\citet{CWY88} calculated upper limits to 
companion masses in the mass-period plane for the stars in their sample. 
While six stars (including $\epsilon$ Eridani) exhibited velocity scatter greater than 
2.5 $\sigma$, no exoplanets were detected in this small sample.

A significant body of work at the time \citep{DLN81, D82, D85, GT85} indicated that 
stellar signals, including spots, p-mode oscillations and magnetically modulated 
granulation changes could introduce line profile variations that might be 
misinterpreted as Doppler shifts.  The spectral format 
used by \citet{CWY88} did not include the Ca II H\&K lines, known 
to be a good diagnostic of chromospheric activity. Instead, they 
monitored the equivalent width of the 8662 \AA\ Ca II infrared triplet line for line 
filling from chromospheric emission. Coherent variability was measured in the 
equivalent widths of the 8662 \AA\ line for some of their target stars, however, no 
correlations were found between this activity metric and the measured radial 
velocities. \citet{CWY88} also implemented an observing strategy 
to deal with p-mode oscillations identified by \citet{N84} in $\epsilon$ Eridani:
to average over these oscillations, the exposure time 
for this star was set to be a multiple of the 10-minute oscillation period.

There were serious challenges with the HF reference cell. The HF lines 
were pressure sensitive and the cell had to be run at the high temperature 
of $100^\circ$C. A relatively long 0.9-m  path length was required to provide a 
sufficient column density of HF molecules and the cell had to be refilled for 
each observing run.  Perhaps most important, HF is a dangerous toxic 
substance.  Other reference cells had also been developed and showed 
equal promise in delivering high precision. Iodine cells were a particularly 
promising choice and provided a wavelength standard that had long 
been used in lock-in amplifiers to stabilize lasers. \citet{B77} used an iodine reference cell 
in the light path of a spectrometer to measure the relative Doppler shift for a 
single Ti II line (5713.9 \AA) in the solar umbra; this technique 
yielded a precision of 15 \ms.  \citet{L88, L89} later used an iodine reference cell to 
measure stellar p-mode oscillations at the Palomar coud{\'e} spectrograph. 

At about the same time, \citet{McMillan86} investigated the use of a Fabry-P{\'e}rot 
interferometer to measure stellar Doppler shifts induced by orbiting planets.
They used optical fibers to couple starlight into an echelle spectrograph to 
minimize the effects of guiding and seeing on the instrumental profile and injected 
a comb of etalon lines with a finesse of 12, calibrated to argon and iron emission lines 
and tuned to compensate for the barycentric velocity at the telescope. Although the 
program did not detect any exoplanets, some of the innovations, in particular 
the scrambling of light with fibers and etalons for wavelength calibration, 
are being revisited with success today. 

\citet{CH90} constructed a prototype ${\rm I_2}$ cell and reported 
that their day sky tests in November and December 1988 at the 2.7-m McDonald 
telescope indicated that the iodine technique might be able to achieve routine 
Doppler measurements with precisions better than 5 \ms\ in stars with sufficient 
photospheric absorption lines.  In a follow-up pilot program on the lower
resolution 2.1-m coud{\'e} spectrograph at the McDonald Observatory, they 
used an iodine reference cell to measure the radial velocity shifts of 
a set of evolved giant stars. The spectral range for these measurements 
was only 23 \AA, centered at 5530 \AA\ and the iodine cell was not temperature 
controlled; at times, iodine was observed to condense out of its gaseous state. 
Although \citet{CH90} commented on the importance of 
modeling the instrumental profile, they did not carry out this step. 
Despite these compromises in the early program, the pilot project achieved 
a velocity precision of 10 - 15 \ms. 

When \citet{MB92} began a Doppler planet search at Lick Observatory in 1987,
they elected to use an approach similar in principal to that of 
\citet{CW79}. They began tests of precise Doppler measurements with 
an iodine reference cell at the entrance slit of the Hamilton spectrograph 
\citep{V87} at Lick Observatory. Particular attention was paid to minimizing 
changes in the illumination of the spectrograph: a collimator mask was used 
to ensure that small movements of the telescope pupil with hour angle would 
not cause shifts in the position of the collimated beam on the downstream optics, 
the thorium argon lamp was focused each night with an algorithm to 
ensured that the same emission lines hit the 
same pixels on the CCD detector each night, and the alignment of the iodine 
cell with the telescope optical axis was checked at the beginning of each run. 
Perhaps the most important breakthrough was the successful modeling of the  
instrumental profile \citep{MB92, VBM95, BM96} also called the point spread function 
(PSF) or most accurately the spectral line spread function SLSF \citep{S13}.  \citet{MB92} 
developed a rigorous Doppler analysis code to model the observed spectrum; free parameters  
included shifts in the stellar lines from barycentric or orbital motion, the 
superimposed iodine lines, and the SLSF from the Hamilton spectrograph. 

The 3 \ms\  breakthrough Doppler precision of the Lick planet search program
allowed for the confirmation of radial velocity signals consistent with 
a gas giant planet orbiting 
51 Peg b \citep{MQ95, M97} and for the detection of dozens of other exoplanets.
In the early days, little time and very modest funding was allocated to the 
project, yet the Lick Planet Search ran for almost twenty five years, from 1987 until 2011. 
For all practical purposes, the project ended when the insulation 
around the iodine cell overheated and damaged the cell. Here, we present 
the radial velocity measurements that were obtained with that program. 

\section{The Lick Planet Search} 
\label{LPS}
The Hamilton spectrograph is a high-resolution optical echelle spectrometer 
that is located at the coud{\'e} focus of the 3-m Shane telescope with a second 
light feed from the Coud{\'e} Auxiliary Telescope (CAT), which is a siderostat-fed 
0.6-m telescope. The CAT enables use of the spectrometer when other instruments 
are being used on the 3-m telescope. The free spectral range can include
3400 \AA\ to 9000 \AA\ in a single observation. The instrument resolution varies from 
50,000 to 115,000 depending primarily on the choice of decker (width of the entrance slit). 

\subsection{Doppler Analysis}
\label{DA}
The Lick planet search began in 1987 June when an iodine cell was installed 
in front of the entrance slit to the the coud{\'e} Hamilton spectrograph
\citep{V87}. The version of the Doppler code used to derive velocities for 
the stars presented here was developed by Paul Butler and Geoff Marcy and 
is described in detail in a series of papers: \citet{MB92, BM96, VBM95}. 
However, the Doppler analysis for more than 14,000 velocity 
measurements presented in this paper was rerun with a single version of 
the Doppler code, on a single computer and one version of IDL to ensure 
version control of subroutines, control of model inputs like the 
SLSF description and the templates and general consistency in the 
analysis before archiving the data.

The iodine absorption cell acts as a transmission 
filter, imprinting thousands of narrow iodine absorption lines on the 
stellar spectrum in the wavelength region from 5010 to 6200 \AA. The iodine 
lines provide a fiducial wavelength scale and are used to model the 
wavelength-dependent SLSF of the echelle spectrometer. 
Because the SLSF changes over the echelle format of the 
CCD detector, the spectrum is broken into 2 \AA\
(40-pixel) chunks from 5010 to 5800 \AA\ (orders with telluric contamination
were excluded in this version of the Doppler code, but are more surgically masked 
out in later versions). 

The Doppler analysis is carried out for each of these 
chunks independently with a forward modeling technique. The Doppler 
model requires the true intrinsic stellar spectrum (ISS) for each of the 
target stars. The precursor to the ISS is a ``template" spectrum of the program star 
obtained with high resolution and high signal-to-noise ratio (SNR) without the iodine 
cell.  Because all spectra are broadened (often by an asymmetric 
SLSF), the template spectrum is deconvolved using a SLSF that is itself a 
model (see Section 3.2) that maps the high-resolution FTS spectrum of the cell to an echelle spectrum 
of the iodine cell (illuminated by a featureless B star). 
The B-star iodine cell observations bracket the template spectrum in time to 
minimize changes in the SLSF.  One ISS is generated for every program 
star on the Lick Planet Search project. 
The ISS is then multiplied by an appropriate segment of the 
high-resolution FTS spectrum of the iodine cell and this product is convolved 
with a new SLSF model to fit the time-series spectra obtained with the iodine
cell. The modeling process is driven by a Levenberg-Marquardt 
algorithm and the free parameters for each 40-pixel chunk include 
the wavelength zero point, dispersion, continuum normalization, 
a Doppler shift and eleven free parameters that describe the SLSF. 
Each chunk has its own SLSF model, wavelength solution 
and Doppler shift and differential velocities are calculated independently for 
each chunk over time.  The nightly radial velocity is calculated as the 
weighted mean of the Doppler shifts for all of the chunks and the 
uncertainty is the standard deviation of the chunk velocities. 

\section{Stability Challenges} 
Over the twenty-five year time span of the Lick planet search there were  
many changes that were made to support the planet search project and 
the general user community. The instrumental profile was very asymmetric and 
the instrument resolution was ${\rm R \sim 40,000}$ until the optics of the 
Schmidt camera of the spectrometer were upgraded by installing  a 
new corrector plate and field flattener in 1994 November.  Before the camera upgrade, the 
single measurement Doppler precision for the Lick Planet Search 
was $\sim$10 \ms. After the optics upgrade and an upgrade to a larger CCD detector, 
the Doppler precision improved to about 3 \ms.

Before internet time was available, the current date and
time were manually set each night on the Disk Operating System (DOS) computers 
by phoning to get a recorded message of the time. Before auto guiders were 
implemented, manual guiding of the star on the slit was done with an RA-Dec 
paddle (while viewing crosshairs through a periscope).  The original slit and decker 
consisted of blades in front of a V-shaped jaw so that small variations in the actual 
slit width could occur for mechanical reasons like backlash in the slit motor. 
The slit and decker system was replaced with an aperture plate in 1998 that 
presented a set of machined rectangular slits at the entrance to the spectrometer. 

The Hamilton spectrograph is a facility instrument and the position of the 
spectral format on the detector can be adjusted with a motorized grating tilt or 
a change in the vertical position of the dewar.  The instrument is regularly reconfigured by observers 
with different programs, introducing a significant source of instability for the precision 
Doppler program. To initialize the spectrometer setup, a focus routine is run at the 
beginning of each planet search night to insure that emission lines from the thorium 
argon lamp fall on the same pixel, giving a wavelength solution that is constant to 
roughly $\pm 0.5$ pixel. 

Although the instrument is in a closed room at the coud{\'e} focus, the south-facing 
wall of the coud{\'e} room is warmed by direct sunlight. A weather station that 
monitored temperatures in the coud{\'e} room in 2001 recorded typical 
diurnal swings on the Celsius scale of a few degrees and seasonal 
changes of about ten degrees. The Hamilton spectrograph optics were mounted 
on an I-beam in the coud{\'e} room and would have experienced some thermal 
expansion and contraction. There was also significant variability in 
pressure and humidity. A dehumidifier operated during
the summer months. In the winter, a stream of ${\rm N_2}$ gas flowed 
across the dewar window to mitigate condensation. 

Precise corrections for the barycentric motion of the Earth are critical for 
accurate Doppler measurements. An error of 30 seconds in the exposure midpoint time 
contributes up to 1 \ms\ error in the measured stellar velocity. The 
geometrical midpoint of an exposure is only the right value to use if the 
conditions are perfectly clear.  If the seeing is variable or if there is variable cloud coverage, 
then a photon-weighted midpoint should be used for calculating the barycentric 
velocity. To calculate the flux-weighted midpoint time,  
a photon integrator was used that predated the Hamilton spectrograph. 
The photon integrator consists of a propeller blade behind the slit 
that rotates at a few Hz and diverts about 5\% of the light to a photomultiplier 
tube \citep{K06}. A running photon count is displayed in the software interface 
so that the exposures could be terminated when the desired SNR was reached;
this allows for better uniformity in the quality of spectra. An upgrade to the 
exposure meter allows for automatic termination of observations when a specified SNR is acquired. 
Even with an exposure-meter, barycentric errors can accumulate for subtle reasons,
including small non-linear terms in the barycentric motion or sampling errors in the 
photon counter. To further limit errors in the photon-weighted midpoint time, exposure times are limited to a maximum of 20 minutes. If higher SNR is desired, velocities are determined for each observation 
and consecutive observations are averaged. 

\subsection{CCD changes}
One of the most impactful changes to the Lick planet search project involved 
a series of upgrades of CCD detectors. The pixel size and charge transfer 
characteristics of the detector affect the spectral resolution and 
contribute to the SLSF so that CCD replacements can introduce velocity offsets 
that required zero-point corrections based on "standard" stars on the program 
(stars with low rms velocities). The initial commissioning of the Lick planet 
search in 1987 used a TI $800 \times 800$ pixel CCD. This detector was 
upgraded in 1990 to a $2048 \times 2048$ Ford Aerospace CCD detector,
and the spectral format spanned 4850 - 6700 \AA. Only the wavelength 
segments containing iodine (5090 - 5787 \AA) were used in the Doppler 
analysis, including slightly more than 300 chunks that were 40 pixels wide. 
None of these pre-upgrade observations were reanalyzed for this paper; they 
have been archived since 1997 with a zero-point velocity locked to data obtained 
with the replacement CCD for this project (Dewar 13, discussed below).

The star $\tau$ Ceti was observed to have low rms scatter of 9.4 \ms\ 
by \citet{CWY88} and was regularly used as one of the radial velocities standards 
for the Lick planet search program. A total of 153 observations of $\tau$ Ceti
were made with these first two detectors at Lick. The observations spanned 7 years 
from September 8, 1987 to Sept 27, 1994. The median error bar 
was 9.6 \ms\ with an unbinned velocity RMS of 12.3 \ms. 
During this so-called pre-upgrade era at Lick (i.e., before the Schmidt camera 
optics were refurbished) nightly velocity corrections were made at the level 
of 10 \ms\ to account for zero-point errors from uncontrolled changes in 
the instrumental setup. 

\subsection{Dewar 13}
In 1994 March (just before the Schmidt camera was upgraded in 1994 November), 
the Hamilton Spectrograph detector was upgraded and mounted in Dewar 13. By 
custom, the CCD is assigned the name of the dewar in which it is mounted. The detector 
specifications for this and all subsequent detectors are archived on the Mount Hamilton 
website\footnote{http://mthamilton.ucolick.org/techdocs/detectors}. 
The CCD in Dewar 13 was a $2048 \times 2048$ ``Lick3" device with 
15-micron pixels and a quantum efficiency (QE) of about 23\% over the iodine 
region. The full spectral format spanned from 4850 - 9362 \AA\ and there were 
704 chunks in the iodine region, spanning 5090 to 5787 \AA. Using Dewar 13, 
94 observations of $\tau$ Ceti were obtained  in the 3 years from 1994 November 18  
to 1997 December 12. The median error dropped from 9.6 \ms\ in the pre-upgrade era 
to 2.4 \ms\ with an RMS of 5.84 \ms. The complete time series velocities 
for $\tau$ Ceti are plotted in Figure 1. 

\begin{figure}
\epsscale{0.7}
\plotone{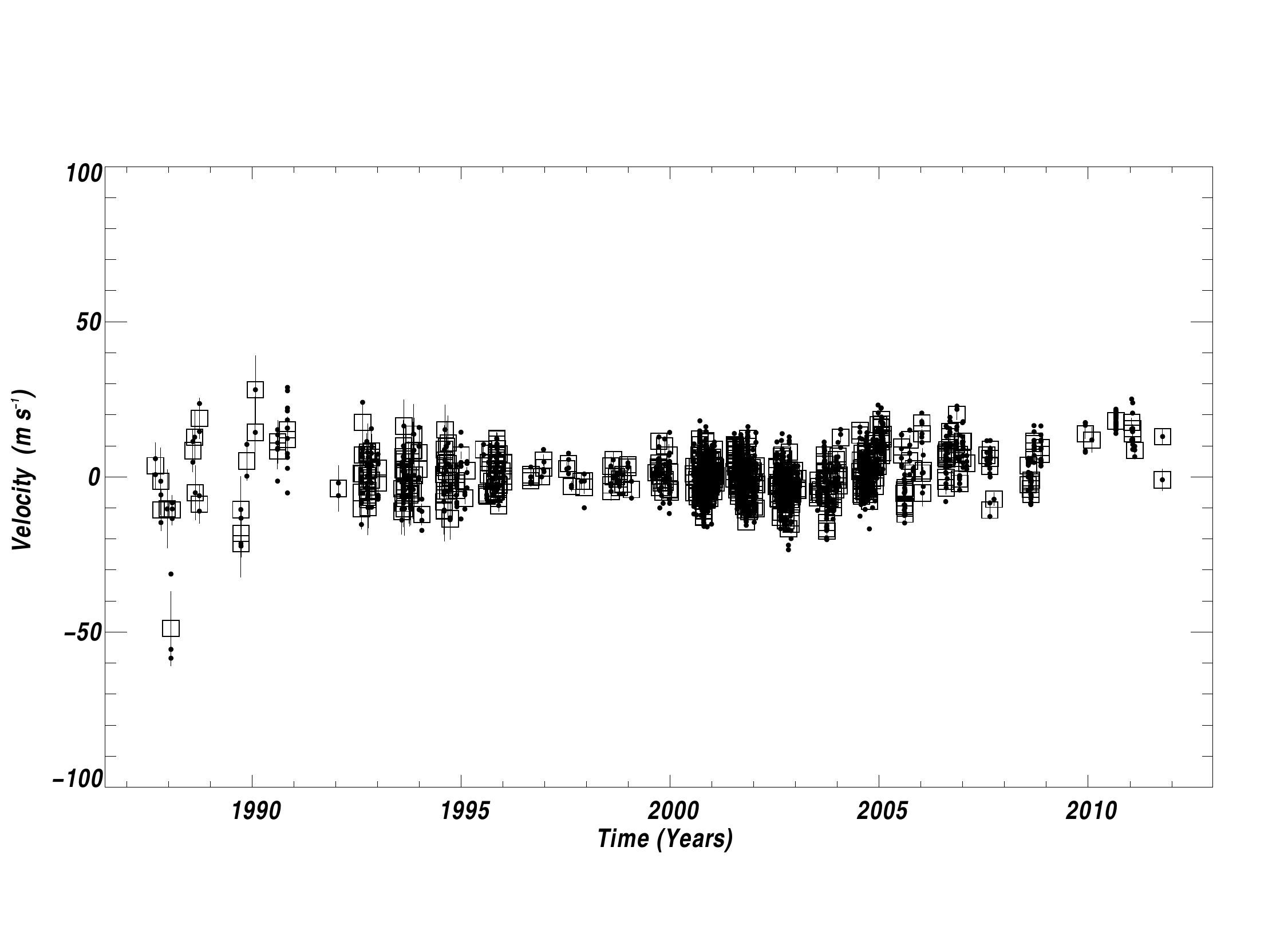}
\caption{Twenty five years of Doppler measurements for $\tau$ Ceti from Lick Observatory. 
The RMS for this complete data set is 7.9 \ms\ and the mean error is 1.7 \ms. This represents 
an example of low RMS scatter for a star on the Lick Planet Search.}
\end{figure}

\begin{figure}
\epsscale{.7}
\plotone{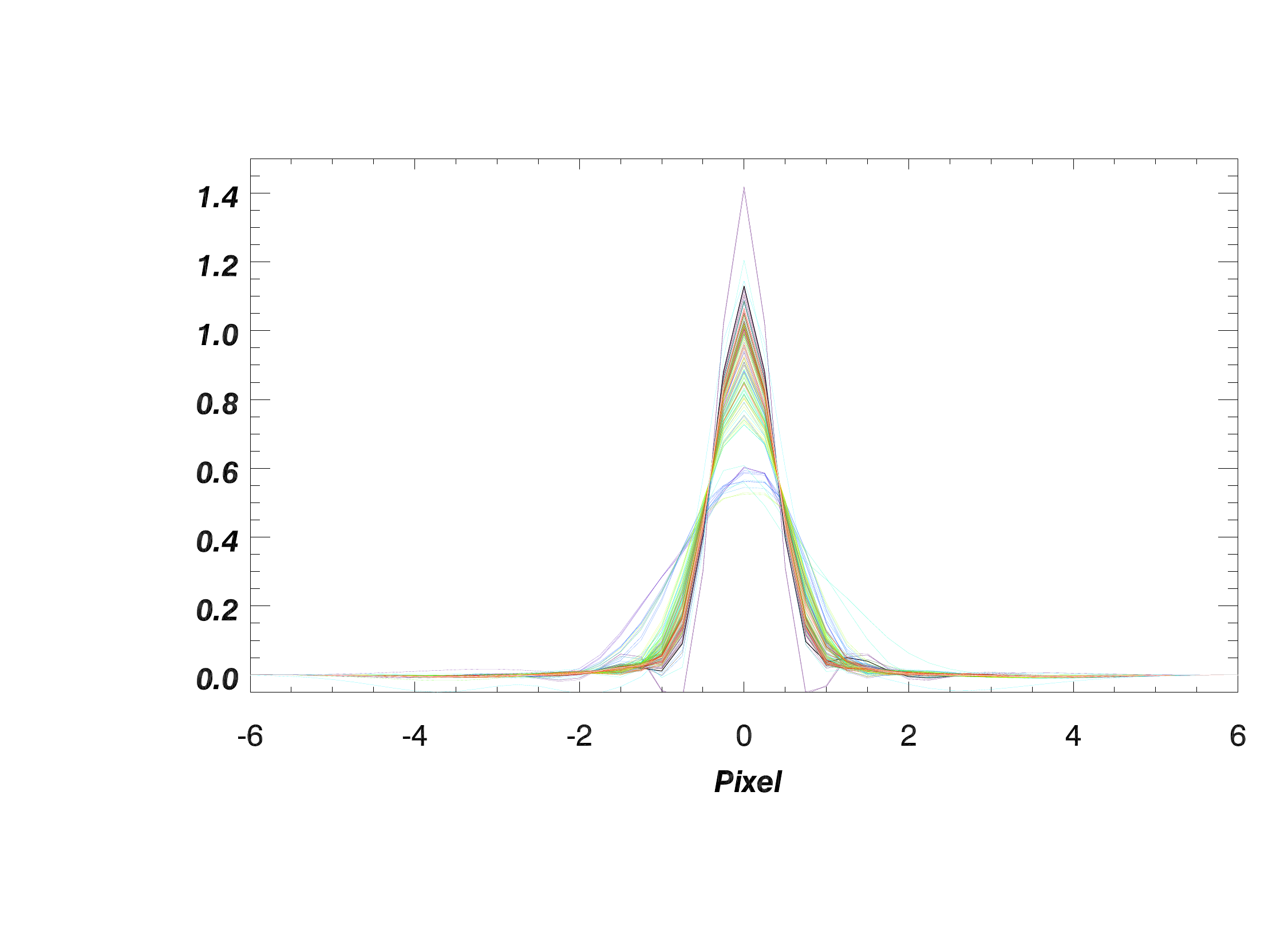}
\caption{SLSF models are over plotted for a single central chunk in the iodine wavelength region.
The models were fit to observations of all B stars observed between 1995 and 1997 
using the CCD in Dewar 13 on the Hamilton spectrograph. Two families of modeled SLSF result from using
the CAT (narrower) and the Shane 3-m (broader).
Considerable variation in the SLSF is apparent, typical of slit-fed spectrometers. }
\end{figure}

Two years of SLSF models for a particular central chunk in the iodine region of Dewar 13 
observations are shown in Figure 2.  These SLSF models were derived for spectra of featureless 
B stars observed from 1995 to 1997 with Dewar 13 at Lick Observatory. Because the 
B stars are inherently featureless and follow the same light path as our program stars, they simply serve to 
illuminate the iodine cell. Thus, the models in Figure 2 show the smearing function (SLSF) that transforms a 
high resolution FTS spectrum of iodine (Figure 3a) into our observed spectrum (Figure 3b). 
The taller narrower set of SLSF models in Figure 2 are fitted to spectra observed with the CAT 0.6-m
telescope and the shorter and broader family of SLSF models are for spectra 
obtained with the 3-m Shane telescope.  
The Shane telescope and the CAT both have f/36 focal ratios, however the plate scale 
is 1.89 arcseconds per millimeter for the Shane and 9.45 arcseconds per millimeter for 
the CAT. This difference in the image magnification at the slit results in the different 
spectral line spread function models in the observed spectra.

\begin{figure}
\epsscale{1.1}
\plottwo{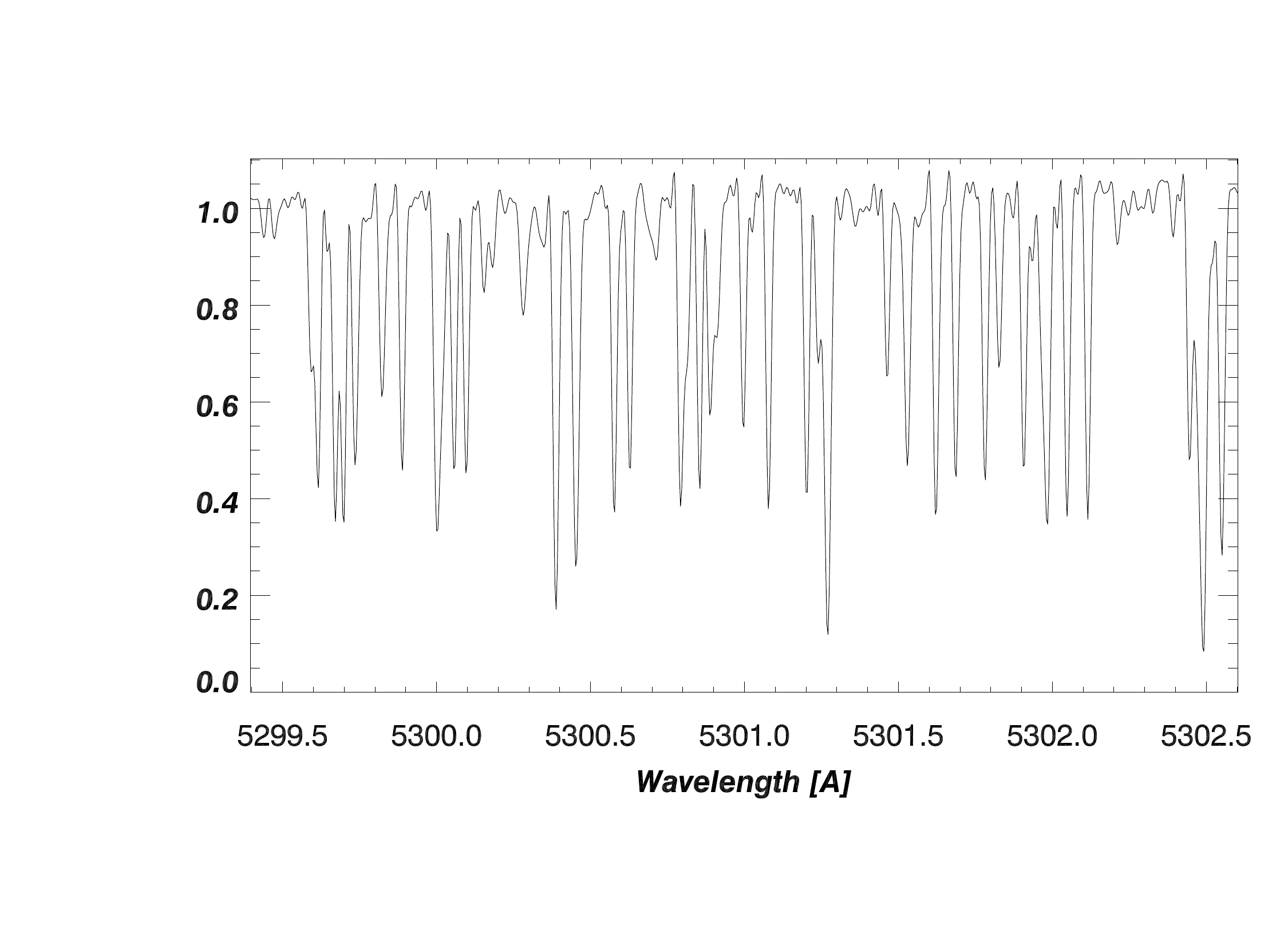}{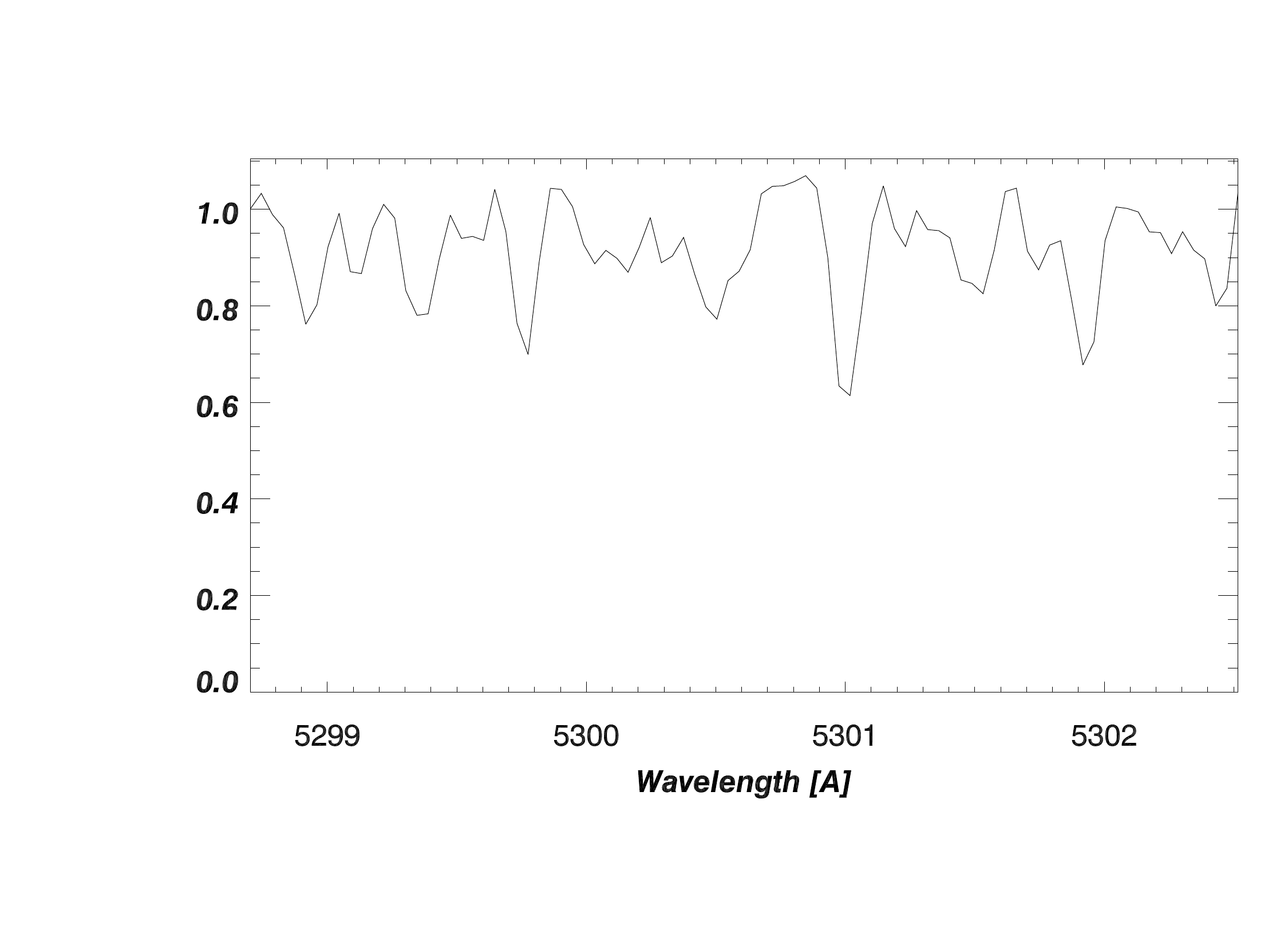}
\caption{(left) the FTS high resolution spectrum of iodine is plotted with flux as a 
function of vacuum wavelengths and (right) the 
observed spectrum for that same wavelength segment when observed 
with the Hamilton spectrograph using the CCD in Dewar 13 (plotted with a wavelength solution
derived from Thorium-Argon lamp calibration. The observed spectrum on the right is 
a convolution of the FTS spectrum with the SLSF shown in Figure 2.}
\end{figure}

In addition to the two families of SLSF models in Figure 2 (CAT observations vs 
those from the 3-m Shane telescope), it is 
clear that there is substantial variation in the SLSF. Variability in the SLSF 
can be caused by changes in temperature and pressure in the spectrograph, 
guiding errors, differences in the motor-driven slit width or variability in seeing. The 
variability that is so apparent in the SLSF demonstrates that the Hamilton optics 
are not being illuminated in the same way. If the SLSF model is not 
accurate this is a source of instability that contributes to systematic errors in the 
Doppler measurements. Errors in the SLSF model can be introduced by cross-talk 
with the wavelength solution or by attempts to minimize $\chi^2$ by fitting noise. 

A key advantage of the Dewar 13 CCD was the relatively narrow SLSF.  The median 
resolution measured from thorium argon observations with the 1\farcs 2 slit 
was $R = 64,000$. However, a significant disadvantage to the Dewar 13 detector 
was the low quantum efficiency. 

\subsection{Dewar 6}
In February 1998, we upgraded to a ``Lick-3" $2048 \times 2048$ CCD 
that had been thinned by Michael Lesser to yield a QE of about 88\% in 
the iodine region. This CCD was also assigned the name of the dewar in which 
it was mounted, Dewar 6. Unfortunately, the Dewar 6 device had a 
slightly concave shape and suffered from charge diffusion so that the SLSF 
was both more asymmetric and broader than Dewar 13.  Four years of SLSF models 
fitted to B star illuminated iodine cell observations are shown for a 
particular wavelength chunk in Figure 4. The broader SLSF resulted in a 
drop in resolution to R=47,000 when observing with the 1\farcs 2 slit. 

\begin{figure}
\epsscale{0.70}
\plotone{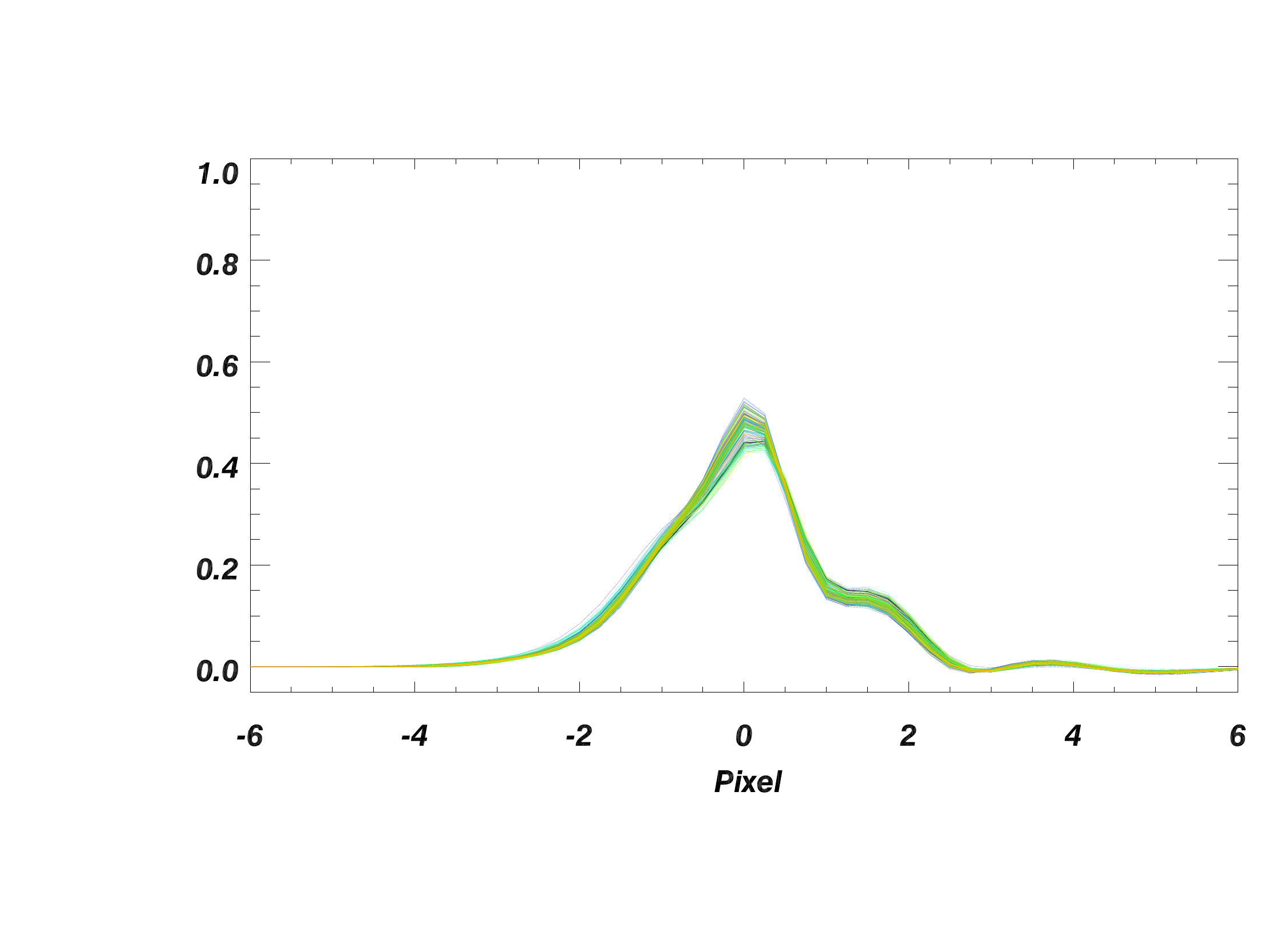}
\caption{SLSF models are overplotted for a single central chunk in the iodine wavelength region.
The models were fit to observations of all B stars observed between 1997 and 2001 
using the CCD in Dewar 6 on the Hamilton spectrograph. }
\end{figure}

We obtained 869 observations of the radial velocity constant star, $\tau$ Ceti, 
with the Dewar 6 detector. Most of the spectra were obtained over a four year period 
from July 30, 1998 to August 29, 2002.  This detector was  used once as an 
engineering back-up for an observing run in Jan 2006. The Doppler analysis was carried out 
for 704 40-pixel chunks spanning the wavelength range 5090 to 5787 \AA.  The median 
internal single 
measurement error was 2.93 \ms\ with a RMS of 6.31 \ms. Thus, this particular detector 
was more efficient, but yielded a slightly poorer Doppler precision. 

\subsection{Dewar 8}
We were reluctant to make another upgrade to the detector because every major change to 
the instrument has a high probability of introducing systematic radial velocity offsets that compromise 
our ability to detect long period planets. However, in 2001, we were given the opportunity 
to try (without cost) a deep depletion device manufactured by LBNL. 
The silicon substrate in this chip was 200 microns thick, so a large voltage across the CCD 
was required to control lateral motion of the electrons (charge diffusion). Upon testing 
we found that we largely recovered the narrow and nearly symmetric SLSF of the 
Dewar 13 CCD, yielding a resolution of 56,000 while providing a QE of 75\% in the 
iodine region.  A representative set of SLSFs for the B star illuminated iodine cell observations taken 
from 2001 to 2011 (again for the same wavelength chunk) are plotted in Figure 5. 
We adopted this new CCD which was mounted into Dewar 8 at Lick Observatory. 
At the time that this detector change was made, we also increased the 
spectral format to include the Ca II H\&K lines  as a chromospheric activity diagnostic. 

\subsection{Dewar zero-point offsets}
\label{dewaroffsets}
Each of the program stars had a unique deconvolved template spectrum, or ISS (see Section 2.1), and that 
same ISS was used to model relative Doppler shifts for the star over the entire time span 
from 1995 to 2011. However, each CCD change can be accompanied by zero-point 
offsets in the derived velocities, in part because of changes to the SLSF. 
To check for zero-point offsets that could be introduced by the use of different detectors, 
we selected all stars with more than 100 observations taken between 1995 and 2011 and 
a velocity RMS less than 20 \ms. We further required that at least 10 Doppler measurements were 
available with each of the three detectors (Dewar 13, Dewar 8, Dewar 6). Eighteen stars 
met these selection criteria; the typical number of observations was 30 for Dewar 13 and 
Dewar 6 and 90 observations for Dewar 8. We calculated the median velocity measured in each 
detector era for these ``standard" stars to see if there were significant zero-point offsets.  
We found that the set of Dewar 13 velocities were on average only 0.37 \ms\ higher 
than the Dewar 6 velocities. However the mean Dewar 8 velocities were on average 13.1 \ms\ 
less than the Dewar 6 velocities. 

A reality check comes from the extensive set of measurements for $\tau$ Ceti, 
which was one of the stars in the set of 18 ``standards." 
We obtained 94 observations of $\tau$ Ceti with Dewar 13; 869 observations with Dewar 6; and 
939 observations with Dewar 8.  Consistent with the results for the other standard
stars, we found that the median velocity of Dewar 13 observations were greater by 0.88 \ms\ 
than the median velocity obtained with Dewar 6, and the median of 
the Dewar 8 velocities were 13.6 \ms\ lower than the median velocities for Dewar 6. 

\begin{figure}
\epsscale{0.70}
\plotone{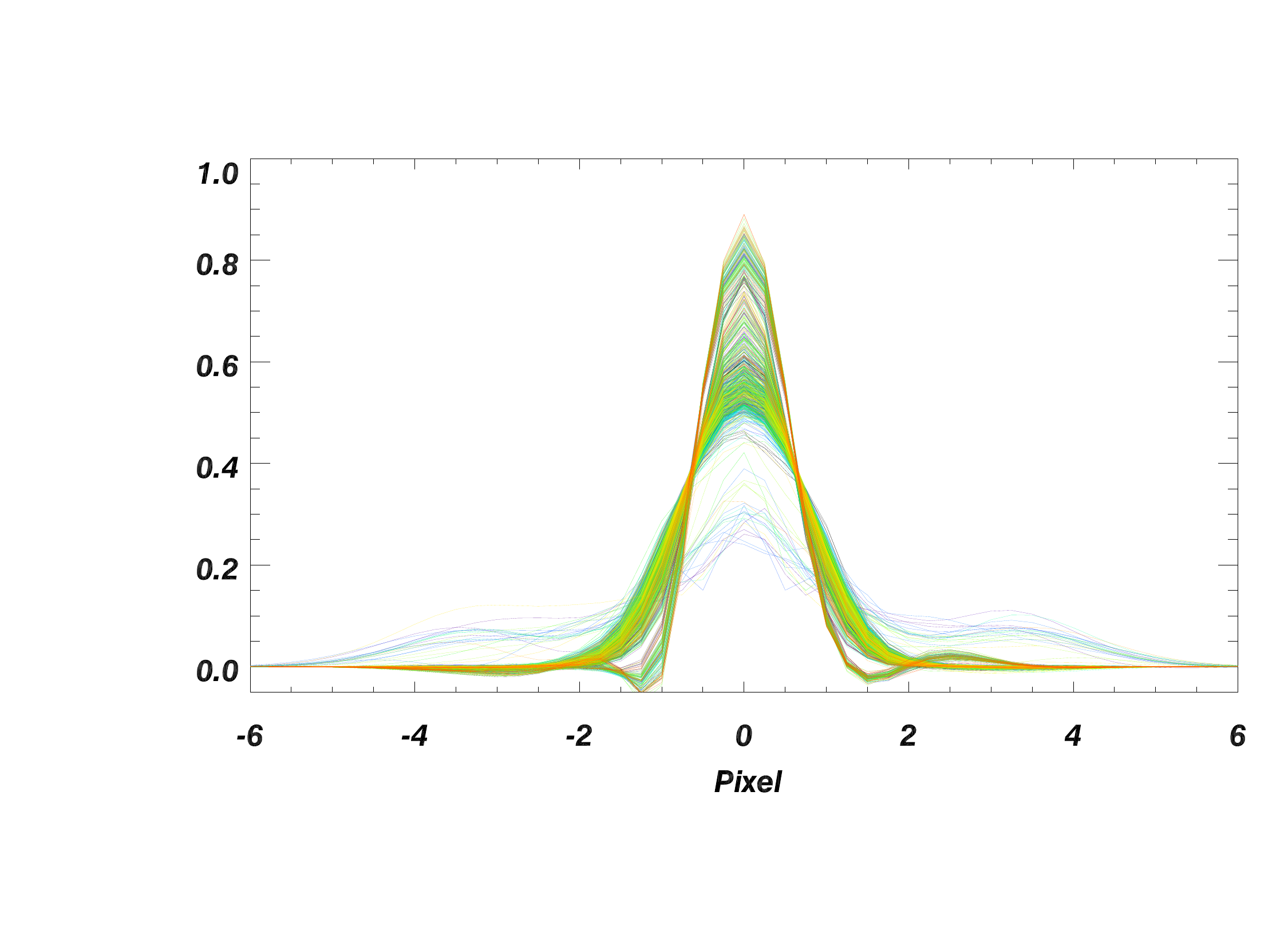}
\caption{SLSF models are over plotted for a single central chunk in the iodine wavelength region.
The models were fit to observations of all B stars observed between 2001 and 2011 
using the CCD in Dewar 8 on the Hamilton spectrograph. }
\end{figure}

Since the calculated offset between velocities obtained with Dewar 13 and Dewar 6 is well below 
our measurement error (-0.37 \ms\ for the set of 18 radial velocity standard 
stars and +0.88 \ms\ for $\tau$ Ceti) we have 
not made any corrections to the Dewar 13 velocities. However, 
we adopted a zero-point correction of +13.1 \ms, added to Dewar 8 velocities for all of the 
Lick planet search stars.  We note that the existence of zero-point 
offsets and our best effort corrections represent systematic errors that can
compromise our ability to detect low-amplitude, long-period exoplanets in the Lick data. 

The high resistivity CCD in Dewar 8 introduced some other problems. The 200-micron silicon 
wafer presented a wider cross-section for interactions with cosmic-ray muons; instead of  
pinpoint cosmic ray events, longer tracks were observed. In addition, the detector is sensitive 
to ambient gamma radiation, which produces Compton scattered electrons that 
random walk in wormlike trails through the detector \citep{Smith02}.  Sources for the gamma radiation 
include concrete, bedrock, and even the BK7 glass in dewar windows. In an attempt to reduce the 
gamma ray flux at the detector the dewar was wrapped with lead shielding. Unfortunately, the lead
shielding turned out to be a source of gamma radiation and the number of recoiling electron tracks 
actually increased. Other shielding materials were tested without a significant improvement, so that 
in the end, we used the dewar without shielding. As a result the low intensity electron tracks had 
the effect of slightly degrading the SNR. Since the number of Compton scattering events scaled 
as the exposure time, the quality of spectra was most affected for faint stars or for the typically 
longer (i.e., 20-minute) observations on the CAT. There were also minor problems with chuck patterns 
on the substrate of the detector that did not quite flat field out, drifts in voltage that changed the 
width of the SLSF, and the thicker detector showed up to $20$\% fringing, increasing for orders 
redward of the iodine region. The mean single measurement (unbinned) error for $\tau$ Ceti 
from the Dewar 8 data was 2.65 \ms\ and the RMS over the ten year period from 2001 to 2011 
was 7.56 \ms. Scatter of 7 \ms\ was typical for long time series Doppler measurements of 
standard stars at Lick and was not considered to be indicative of intrinsic
dynamical velocity variations in the $\tau$ Ceti data set.  

\section{Radial Velocity measurements for the Lick Sample} 
The stellar sample initially consisted of 120 bright FGK and M dwarfs, drawn from 
the Bright Star Catalogue \citep{HJ82} and the Gliese-Jahreiss catalog \citep{G69, GJ79}. 
Binary stars with separations smaller than a few arc seconds were excluded from 
the sample because the flux contamination from a second star would complicate 
the Doppler modeling. In 1997, many of the fainter stars were moved to the 
Keck program and roughly 200 new stars were added at Lick. Since most of the stars
that were moved to the Keck program had \bv\ $> 0.8$, the Lick planet survey was dominated 
by brighter and hotter stars. This can be seen in Figure 6 where the \bv\ distribution 
is plotted. Most of the stars in the Lick sample had 0.4 $<$ \bv\ $<$ 0.7. These 
earlier spectral types contributed to the Doppler errors since the late F and 
early G dwarfs have more intrinsic stellar variability \citep{IF10} and tend to be more 
rapidly rotating.

Motivated by the emerging correlation between stellar metallicity and the formation of 
gas giant planets, \citep{G97, S04, FV05}, an additional 60 metal rich stars were 
added to the program in 2001, bringing the total sample size to
387 stars \citep{F99}. As discussed by \citet{FV05}, adding metal-rich stars 
to the sample does not induce a planet-metallicity correlation because the 
fraction of stars with planets is calculated for each metallicity bin. Rather, 
the Poisson error bars are reduced for any metallicity bins containing a larger numbers of stars.

\begin{figure}
\epsscale{0.70}
\plotone{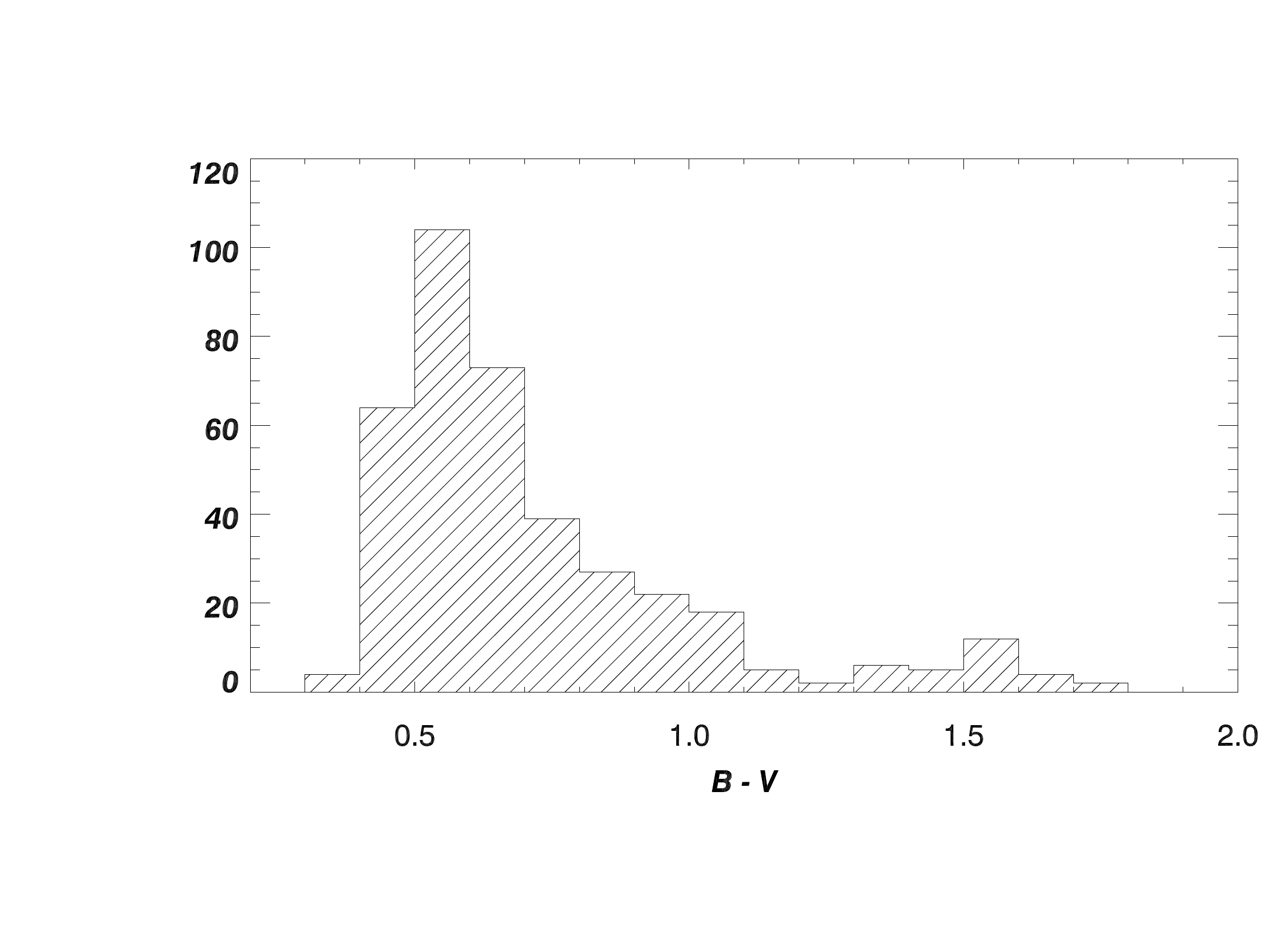}
\caption{The distribution of B-V colors for stars on the Lick planet search sample shows
a selection bias toward brighter and relatively bluer stars.}
\end{figure}

Two other planet surveys were also carried out at Lick: a search for planets orbiting 
nearby K giants \citep{FQ01}, which grew out of a study for the Space Interferometry 
Mission, and a Ph.D. thesis survey for planets around subgiants \citep{J06}; this 
paper does not include the data from those two projects. 

\begin{deluxetable}{llrrrcrrrr}
\tablenum{1}
\tablewidth{0pt}
\tabletypesize{\footnotesize}
\tablecaption{Star sample}
\tablehead{\colhead{  Star ID} & \colhead{RA } & \colhead{Dec} & \colhead{Vmag}
       & \colhead{B-V} & \colhead{Spectral} & \colhead{Nobs}  &  \colhead{Time obs} 
       & \colhead{Median Err}  & \colhead{RV RMS} \\
         \colhead{  ID} & \colhead{(2000)} & \colhead{(2000)} & \colhead{ }
       & \colhead{ } & \colhead{Type} & \colhead{ }  &  \colhead{[Yrs]} & \colhead{[\ms]}
       & \colhead{[\ms]} }

\startdata
         166  &   00 06 36.78  &   +29 01 17  &   6.13  &  0.750  &    K0  &   57  &   24.33  &     7.73  &      21.57 \\
         400  &   00 08 40.93  &   +36 37 37  &   6.22  &  0.456  &    F8  &   19  &   13.19  &     6.86  &      24.67 \\
         984  &   00 14 10.25  &   -07 11 56  &   7.33  &  0.480  &    F5  &    3  &    1.14  &    19.66  &     210.62 \\
        gl14  &   00 17 06.37  &   +40 56 53  &   8.94  &  1.420  &    K7  &   15  &   10.25  &    12.56  &      16.04 \\
       gl15a  &   00 18 22.88  &   +44 01 22  &   8.07  &  1.560  &  M1.5  &   25  &    9.88  &    13.73  &      20.21 \\
\enddata
\tablecomments{This is a stub table with the first few entries of the 
machine readable table (MRT) that lists parameters for 386 stars. 
The MRT is available with the electronic version of this paper}
\end{deluxetable}

Table 1 lists the stars on the long-term Lick planet search program. In the first column 
the star identifier is given. The next two columns contain the right ascension and
declination for epoch 2000. The fourth column contains the visual magnitude of 
the stars. The fifth column lists spectral types that are drawn from the Hipparcos 
catalog \citep{esa97} when available, or listed on SIMBAD. The last three columns list the number of 
observations, the time interval of the observations in years, the median error 
for the data set, and the velocity RMS. The print version of the paper contains a 
stub table listing the first few entries of the complete table that is available online 
in machine readable format. Figure 7 shows the distribution of both the single 
measurement errors (tilted cross-hatched part of the histogram) and the 
RMS velocity scatter (vertical lined histogram) for the full time 
baseline for each star. In some cases, large RMS scatter occurs because 
the stars have exoplanets or stellar companions (no trends or Keplerian 
models were removed when calculating the velocity RMS for Table 1 or Figure 7). 

\begin{figure}
\epsscale{0.70}
\plotone{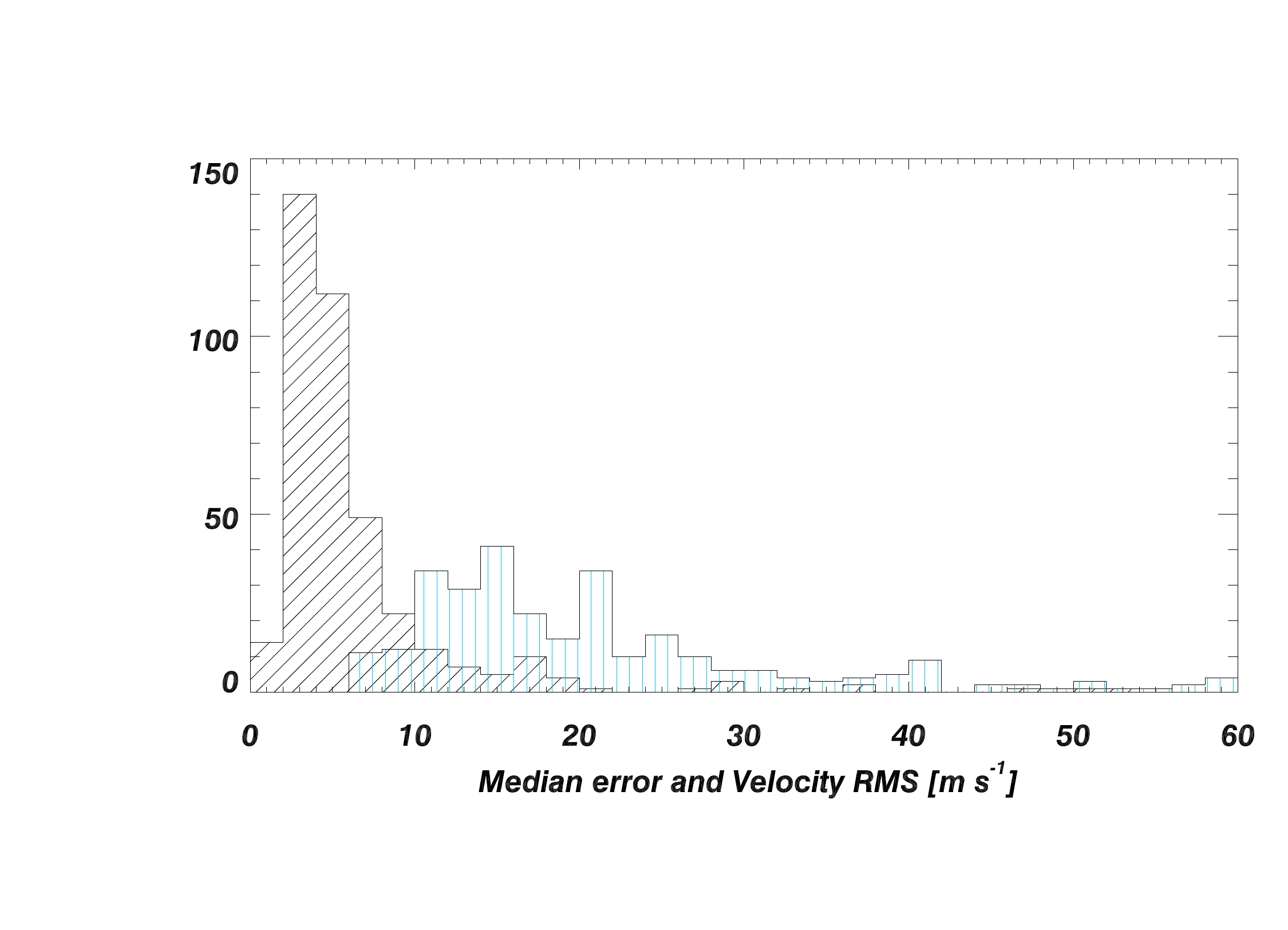}
\caption{The distribution of Doppler errors (angled, cross-hatched histogram) 
peaks at 2 - 4 \ms.  However, most stars exhibit RMS radial velocity scatter 
(vertical-lined histogram) of tens of 
meters per second. This is because the sample includes some stars with 
exoplanets, a large fraction of late F or early G type stars that are 
rapidly rotating and have significant stellar jitter (astrophysical noise sources) 
and because of uncontrolled systematic errors.}
\end{figure}

\begin{deluxetable}{lrrrrc}
\tablenum{2}
\tablewidth{0pt}
\tabletypesize{\small}
\tablecaption{Table of Radial Velocity Measurements}
\tablehead{\colhead{  Star } & \colhead{JD - 2440000.} 
       & \colhead{RV} & \colhead{Error}
       & \colhead{SNR} & \colhead{Dewar} \\
         \colhead{ ID} & \colhead{[days]} 
       & \colhead{[\ms]} & \colhead{[\ms]}
       & \colhead{ } & \colhead{ } }
\startdata
166   &     6959.98020  &    7.95 & 12.65 &  269 &  2 \\
166   &     7046.88100  &    3.05 &  8.95 &  277 &  2 \\
166   &     7373.92230  &   25.78 &  7.73 &  240 &  2 \\
166   &     7846.79660  &   17.41 &  9.13 &  224 &  2 \\
166   &     8113.92840  &    6.84 & 11.68 &  307 &  2 \\
\enddata
\tablecomments{This is a stub table with the first few entries for one star. The 
machine readable table available with the electronic version of this paper lists
more than 14,000 velocity measurements for 386 stars.}
\end{deluxetable}

Table 2 contains the individual radial velocities for the stars listed in Table 1. 
A stub table is shown in the printed version of the journal, but the online paper 
contains the complete table in machine readable format with the star name 
(given as an HD number when available), the observation date in JD - 2440000., 
the measured relative velocity in units of \ms, the formal measurement error in 
units of \ms, the SNR per pixel near blaze 
center of a central iodine order, and the CCD dewar that was used for the observation. 
The zero-point correction discussed in Section 2.6 has been added to the velocities 
obtained with Dewar 8. 

\section{Testbed for Innovation}
\label{innovation}

\subsection{Tip-Tilt at the Coud{\'e} Focus} 

The 3-m Shane telescope has a long focal length and is subject to wind shake. 
On many nights, the image of the star spends less than 50\% of the time on the 
slit. In Fall 1997, we built a prototype tip-tilt system in front of the slit to the 
Hamilton Spectrograph \citep{ME98}. 
The system was mounted on an optical table where a 45-degree fixed position pick-off mirror directed the 
starlight to a tip-tilt mirror mounted on a stepper motor driven gimbal mount. The light was then 
directed back toward the slit by a second 45-degree fixed mirror.  A beam splitter directed about 4\%
of the light to a quadrant photo-multiplier tube (PMT) and a signal from each quadrant was used to 
close the loop by moving the tip-tilt mirror so that all PMT quadrants had equal amounts of light. 
The prototype was tested on three nights 
and demonstrated an improvement in throughput, particularly on one windy night when the open 
loop observations suffered large light losses at the slit. 
However, the RMS scatter of the radial velocities for a set of observations was marginally worse than a 
control case with an open loop. The results were not conclusive because the number
of observations was small, however, one concern is that the tip-tilt mirror was changing 
the angle of light and therefore changing the illumination of the optics in the spectrometer. A better approach
would be to use the tip-tilt system with a fiber scrambler.  

\subsection{Tests with Fiber Scramblers} 
In a series of tests performed with the slit-fed Hamilton spectrograph \citep{S13}, 
we have shown a strong correlation between SLSF variations and changes in star 
position in the sky, most likely due to changes of spectrograph illumination with 
varying telescope positioning (Figure 8; Spronck et al. 2013). 
In addition, we have also documented correlations between SLSF variations and guiding, 
focusing and seeing changes. These tests demonstrated that slit-fed spectrometers are 
unlikely to provide the extraordinary stability required in order to achieve extreme 
radial velocity precision.

\begin{figure}
\epsscale{0.9}
\plotone{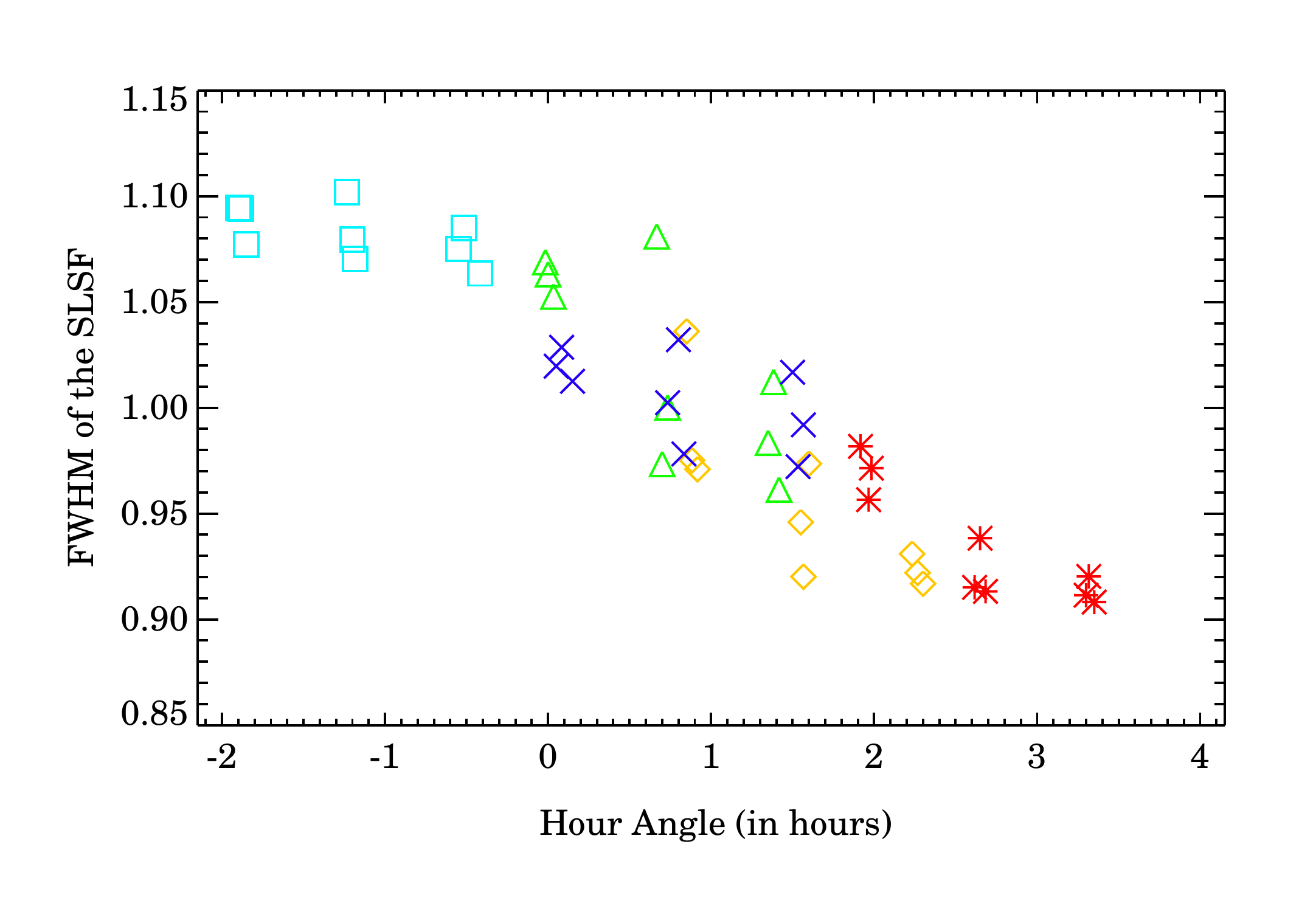}
\caption{SLSF for 45 B star observations as a function of hour angle (reproduced 
Figure 8b from Spronck \etal 2013).
Each colored symbol corresponds to a different B star.}
\end{figure}

In an effort to control short timescale variations in the SLSF, \citet{S13} 
designed a prototype fiber scrambler to feed light to the Hamilton Spectrograph. 
The iodine cell is located about 10 cm in front of the decker plate with space 
for a pickoff mirror to feed the TV guide camera. Because of these space 
considerations and because we wanted to use the existing TV target acquisition 
and guiding system, the fiber scrambler was mounted on an optical breadboard 
behind the entrance aperture to the spectrometer. Tests were made with a 100 micron 
circular fiber that was 15 meters in length and the throughput of the fiber scrambler 
was measured to be 65\%. 

Our analysis of spectra from the B star illuminated iodine cell showed that the variability 
of the SLSF improved by a factor of two when using the fiber (Spronck et al. 2013). 
However, a systematic trend in the full-width half maximum (FWHM) of the SLSF was 
observed as a function of hour angle, even when using the fiber scrambler. 
This suggests incomplete scrambling by the fiber that resulted in output that was 
dependent on the angle that light entered the fiber. 

For test observations of HD 166620 with comparable SNR in both the fiber and the slit 
observations, we found that the velocity RMS decreased by about 30\% for the fiber 
observations. For fiber observations of two other test stars, the velocity was a 
few percent worse. However, the SNR was about 50\% lower for all of the fiber observations 
relative to the slit observations, making a controlled comparison impossible and 
biased against good results for the fiber observations. 

A double scrambler \citep{HR92} was also 
tested with the Hamilton spectrograph. In the double scrambler, the near field image 
of the fiber is inverted into the far field (pupil) that is then injected into a second fiber. 
The output from the second fiber then illuminates the spectrograph. The alignment of 
this system was not optimized and the throughput dropped to 15\%. Yet, the SLSF 
stability improved by another factor of two and the trend in FWHM of the SLSF with 
hour angle that we observed with the single round fiber was 
eliminated with the double scrambler. 

Although the fiber scrambler was only used in engineering mode at Lick Observatory, 
these tests provided valuable information about optimal coupling of light to spectrometers 
for future designs.

\subsection{FTS iodine scans}
In addition to hardware upgrades, we also worked to improve our data analysis techniques. 
The Fourier Transform Spectrograph (FTS) scan of the iodine cell is perhaps the single most important 
component in our Doppler analysis because it affects every stage of the modeling process. 
The FTS iodine spectrum is supposed to provide ``ground truth" for determining the wavelength 
solution, measuring Doppler shifts, and modeling the SLSF. The FTS iodine spectrum is also important 
for generating a second key ingredient in our model: the intrinsic stellar spectrum (ISS). 
As discussed in Section 2.1, the ISS is created by deconvolving a stellar template spectrum, 
a high SNR spectrum without iodine. Since the template spectra do not contain iodine lines, the SLSF used in 
the deconvolution is derived from observations of the iodine cell taken before and after the 
stellar template observations. 

The original FTS spectrum of the Lick iodine cell was obtained at Kitt Peak and had a 
resolution of R $\sim 300 000$.  We obtained FTS scans at the 
Environmental Molecular Sciences Laboratory (EMSL) with R$\sim 1,000,000$ and 
(per resolution element) SNR $\sim 1000$.  This higher resolution FTS scan should provide a more accurate 
SLSF model; however, deconvolution techniques tend to amplify any noise and 
high frequency ringing in the continuum can be seen with some of the deconvolved spectra.  
An ongoing effort is development of deconvolution techniques with wavelet techniques or 
low pass filters to attenuate deconvolved shot noise.   

\section{Loss of the I2 cell} 
\label{loss} 
On 2011 November 11 we learned that the heater for the iodine cell had
malfunctioned, melting most of the foam insulation around the iodine cell. 
The cell itself remained intact, however, a sticky residue dripped over the optical 
windows of the cell (Figure 9). The cell windows were carefully cleaned and 
the insulation was replaced.  

\begin{figure}
\epsscale{0.7}
\plotone{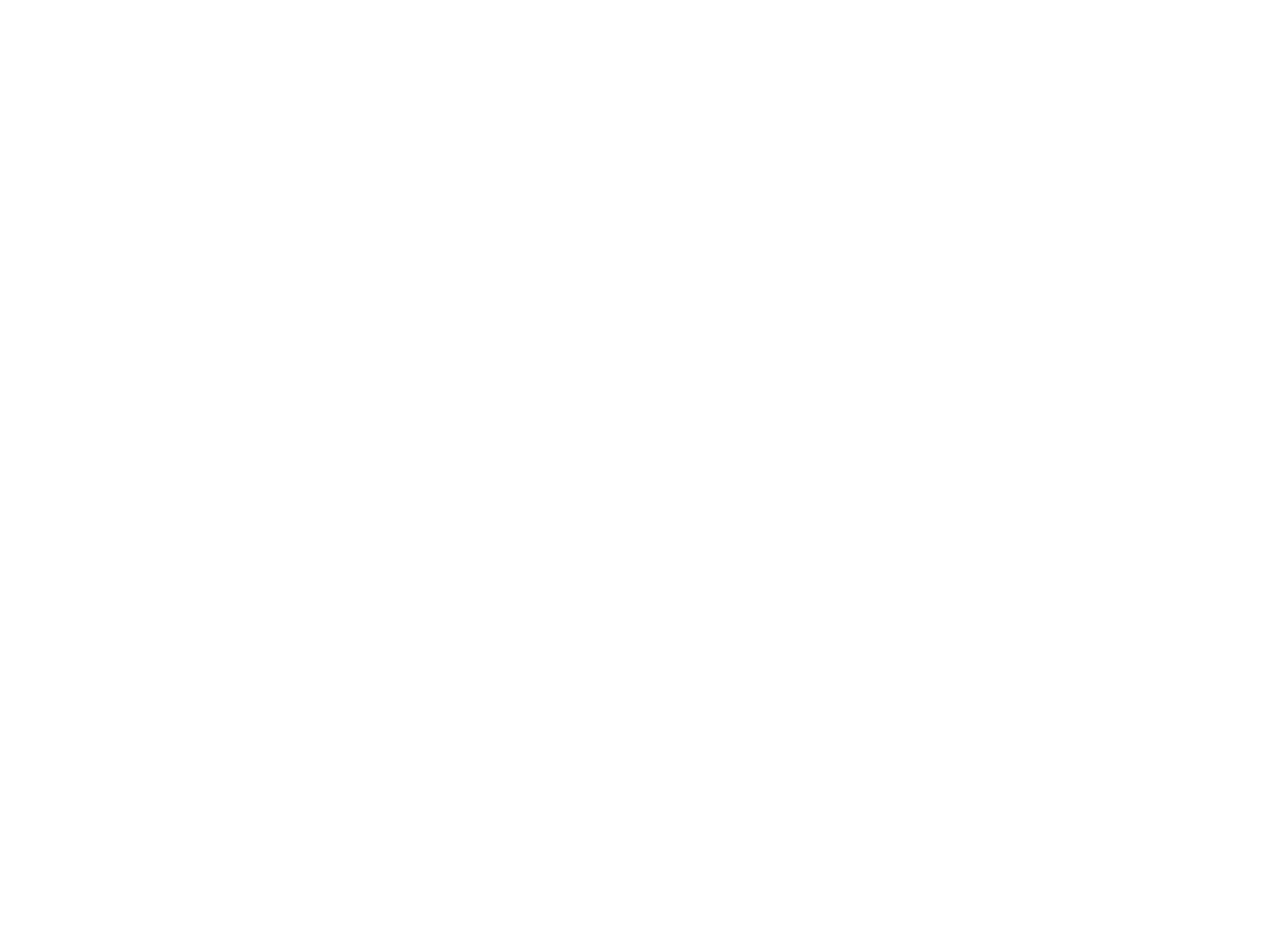}
\caption{Melted insulation dripped across the optical window of the iodine cell
after a heater malfunction in 2011 November.}
\end{figure}

When the first spectra were taken with the refurbished cell, we found that the 
model for the iodine lines yielded an unacceptably poor $\chi^2$ fit.  We visually 
inspected the spectrum and found that the iodine lines from spectra 
obtained after the malfunction did not match those taken before the cell overheated.  
Furthermore, the differences in the spectra were wavelength dependent. 

Figure 10 shows a comparison; an iodine spectrum taken on 
2010 April 26 (``ri62.149") before the cell was damaged is plotted in black and 
an iodine spectrum after the damage, taken in January 2012 (``ri82.49") is plotted in red. 
To highlight the wavelength dependence of the difference, two plots are shown in Figure 10: 
the left plot is a wavelength segment at about 5400 \AA\ and the right plot is a 
wavelength segment at about 5700 \AA. 

Iodine spectra do not normally change and the plotted black spectrum, taken before the cell 
was damaged, is essentially identical to thousands of iodine spectra of the Lick cell 
obtained from 1989 through 2011. However, the plotted red spectrum (from the damaged cell) 
in Figure 10 shows significant changes. 
The spectrum from the damaged cell shows a small wavelength shift at 5400 \AA (Fig 10 left) 
relative to the good iodine spectrum and the 
spectral lines from the damaged cell are deeper and slightly broader.  
Progressing to redder wavelengths, the relative shift in the spectral lines from the damaged 
cell increases as a function of wavelength. 
Figure 10 (right) shows the comparison for a wavelength segment 
at about 5700 \AA: the spectral lines from the damaged cell lines are  
significantly shifted and are now shallower than the good iodine spectrum. 

\begin{figure}
\epsscale{1.0}
\plottwo{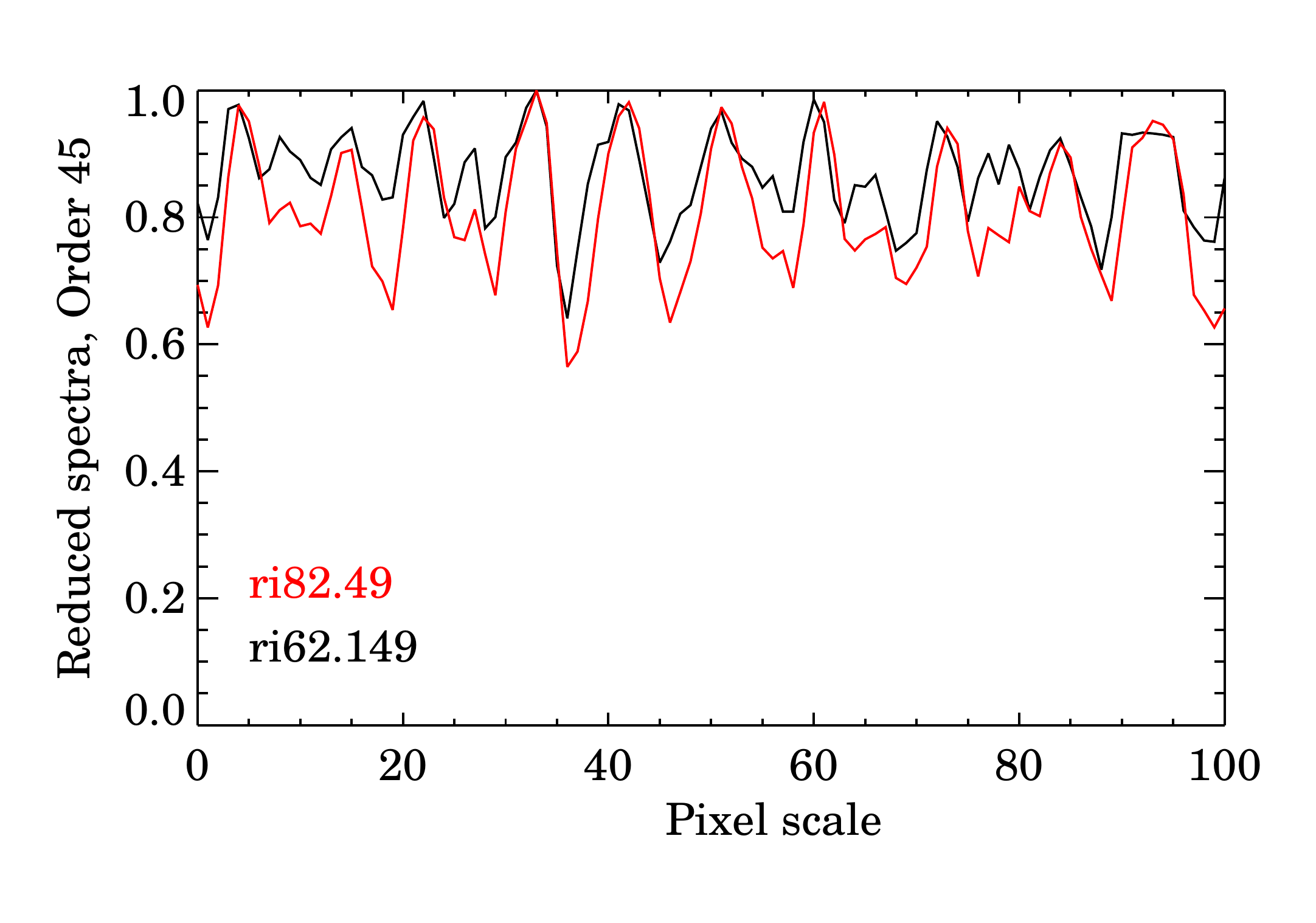}{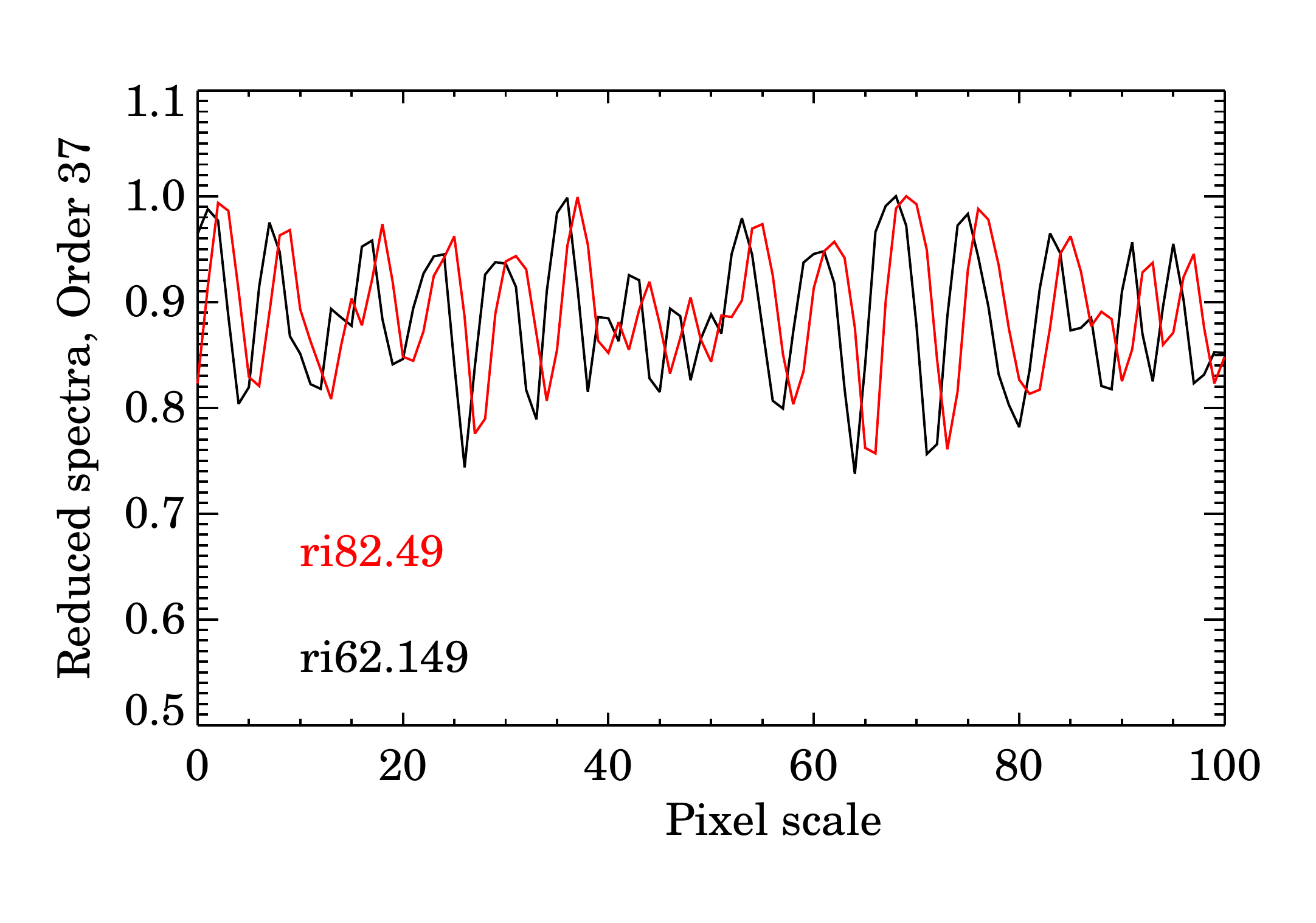}
\caption{A wavelength dependent change was found in the iodine absorption spectrum 
after the iodine cell overheated. The comparison spectrum is shown in black (``ri62.149'') 
and was obtained before the cell was damaged. This spectrum matches iodine spectra 
from the previous two decades and is well modeled with the FTS spectrum of the cell. 
The red spectrum ``ri82.49'' is an iodine spectrum taken after the heater malfunction and is 
plotted in red. (left) The damaged-cell spectrum at about 5400 \AA\ is slightly shifted and 
has deeper and broader lines than the spectrum obtained with the good cell. 
(right) The wavelength segment at about 5700 \AA\ shows a significant 
wavelength shift and the spectral line depths are shallower than the (black) good iodine spectrum. }
\end{figure}

\begin{figure}
\epsscale{0.70}
\plotone{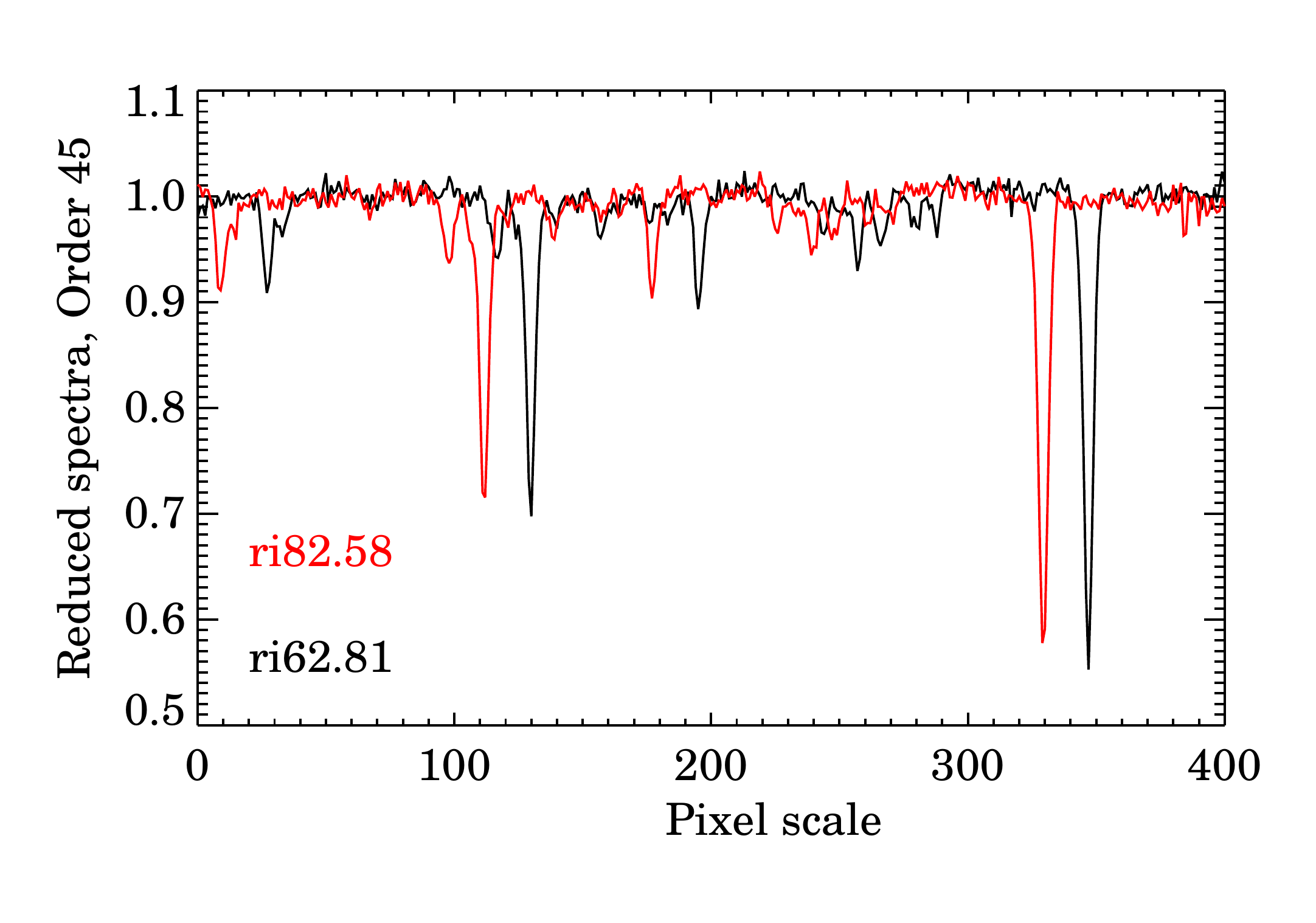}
\caption{A stellar spectrum obtained without the iodine cell on 2010 April 26 (``ri62.81")
is unchanged from a spectrum taken after the cell was damaged (``ri82.58") except that the SLSF 
for the ``ri82.58" observation is slightly broader and therefore the lines are a bit shallower. 
This isolates the observed change in the iodine spectrum to a change in the I2 cell rather than to any 
changes in the spectrometer or detector. }
\end{figure}

This kind of difference in the iodine spectrum was never seen over the many years of the Lick planet search. 
Before the cell overheated, the same iodine segments looked the same at all wavelengths 
year after year.  To demonstrate that this change was not the result of some other problem in the spectrometer, 
we obtained a stellar spectrum (without iodine) and compared it 
to a spectrum of that same star taken a year before. Although there is a difference in the barycentric 
velocity, the stellar spectrum is unchanged (Figure 11). This test isolated the problem to the 
iodine cell.  

We considered several explanations: iodine or other elements might have been 
released from the cell wall during heating.  Wavelength dependent scattering from 
the damaged iodine window might have introduced an apparent change to the 
normalized spectrum. A change in pressure (partial loss of vacuum, 
but not a loss of iodine) in the cell could have occurred during thermal cycling of the cell. 
Whatever the cause, the 
cell was spectroscopically different after the heater malfunction. 

We replaced the iodine cell, but realized that this would require 
a few data points just to calibrate an offset between the old and new 
velocity measurements.  Since we expected that the Automated Planet Finder 
on a new 2.4-m telescope at Lick would soon come online, we elected to 
terminate the program on the Hamilton spectrograph. 

\begin{figure} 
\epsscale{0.80}
\plotone{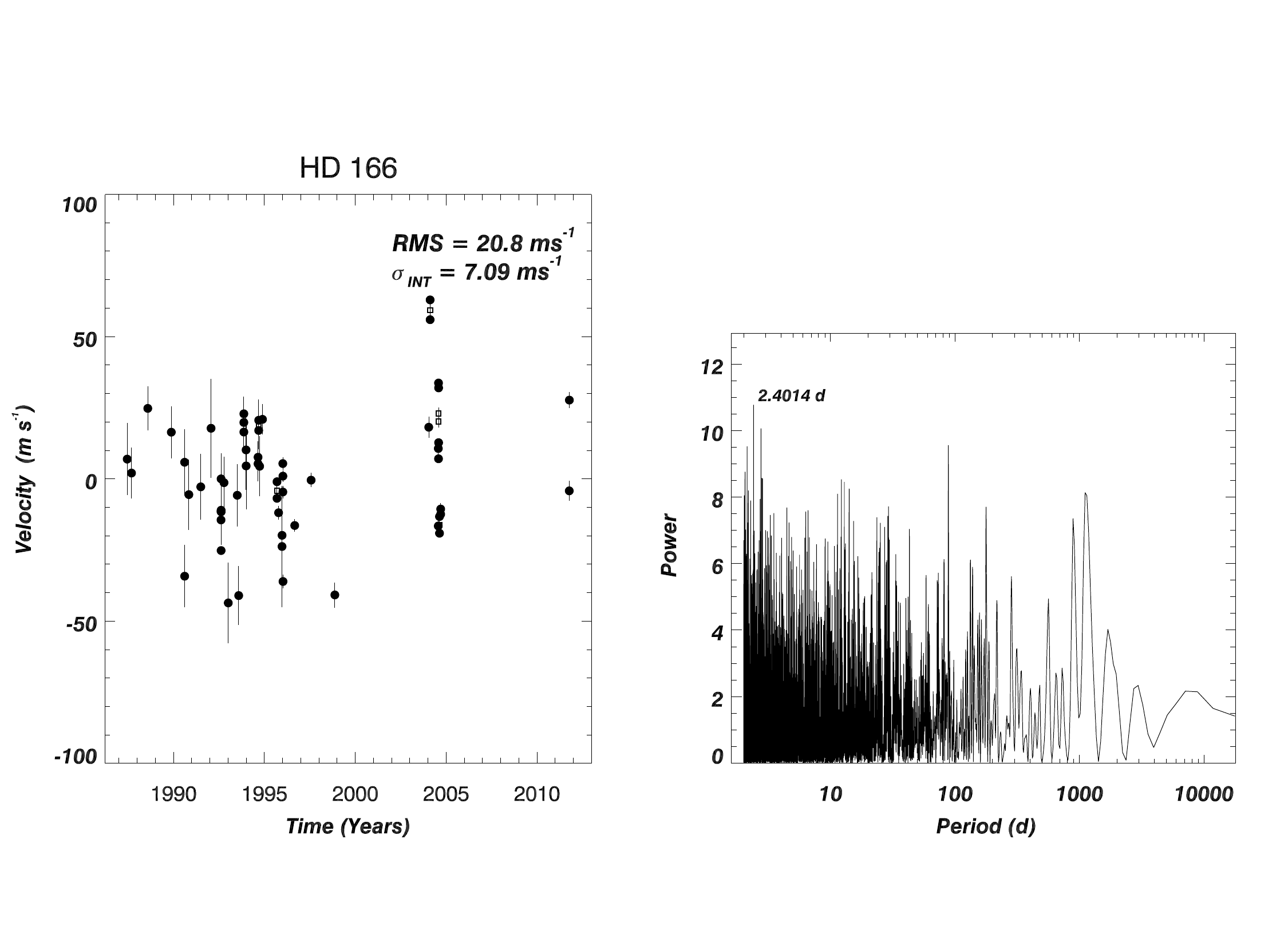}
\caption{Time-series radial velocity data for the star HD~166. This 
is the first example from the online Figure set plotting Doppler velocities for the 386 stars
listed in Table 1. For stars with less than one year of data, the time scale is in Julian days;
otherwise the time is in years. The RMS of the velocities as well as the formal measurement
error are listed in the top right corner of the plot. Stars with more than 12 observations also 
have a periodogram plot included as a companion Figure. }
\end{figure}

\section{Discussion} 
\label{discussion}
The Lick planet search project was a pioneering program 
to measure the reflex velocities induced by exoplanets for a sample of 386 nearby stars. 
The velocity sets are plotted for all 386 stars on the Lick program 
in a figure set with the online edition of this paper.  Figure 12 shows the first 
of these figures for the print version of the paper. 

Many of the first known planets outside our solar system were identified or 
confirmed with the Lick survey, including: 
51 Peg b \citep{MQ95, M97}; the first planet in an eccentric orbit, 70 Vir b \citep{MB96};
47 UMa b, c \citep{BM96, F02a}; $\tau$ Boo b \citep{ BM97}; the 
first planet orbiting a star in a binary system, 16 Cyg B b \citep{C97}; 
the first multi-planet system around  normal star, Ups And b, c, d \citep{BM97, BM99}; 
55 Cnc b, c, d, f, g \citep{BM97, M02, F08}; 
HD195019 HD217107 \citep{F99}; HD 89744 b \citep{K00}; GJ 876 b, c, d \citep{M01, R05}; 
HD 136118 b, HD 50554 b, HD 105252 b \citep{F02b}; one of the first cold sub-Saturn mass planets, 
HD 3651 b \citep{F03a}; HD 49079 b, HD 12661 b, c, HD 38529 b, c \citep{F01, F03b};
HD 30562 b, HD~86264b, HD87883 b, HD 89307 b, HD 148427 b, HD 196885A b \citep{F09}.

Gas giant planets were found in hot and cold orbits, exoplanets were 
detected around stars that were members of binary systems, the first 
multi-planet systems were detected and the project provided data used 
to quantify the planet-metallicity relation. The Lick program also provided 
important phase coverage to expedite exoplanet detections at Keck Observatory. 
At Lick Observatory, we first learned that planet formation is both a robust and 
chaotic process.  

The project served both as a workhorse Doppler survey and as a testbed for 
innovation. Thanks to the generous allocation of telescope time by the UC TAC,
we were able to acquire the first high cadence Doppler data sets and learned that intense 
sampling was critical for unraveling complex systems with low amplitude and multiple planet 
signals. We tested a tip-tilt system at the Coud{\'e} focus of the telescope and found 
that while this improved the throughput, particularly on windy nights, the radial velocity 
precision was slightly degraded - probably because of the changing input angle for 
light into the Hamilton spectrograph. We also tested fiber scramblers at Lick and 
achieved significant improvement in the stability of the SLSF. While our velocity 
precision did not improve significantly with the fiber scrambler, we believe that the 
error budget was dominated by other factors such as the decreased SNR and slight loss 
of resolution with the fiber and with instability in the Hamilton spectrograph itself. 

We monitored, but were unable to control, 
large changes in temperature and pressure and significant changes in the SLSF 
over time for the Hamilton spectrometer.  Over the 25-year history of the program, we 
used several different CCD detectors and learned that despite the poor 22\% quantum 
efficiency of the detector used in 1995 - 1997, the low charge diffusion in that detector 
resulted in a narrower SLSF and higher Doppler precision. The program was only able to 
survive all of these changes with what was then state-of-the-art Doppler precision 
because we used an iodine cell to model the wavelength solution, the Doppler shift,
and the SLSF of the instrument.  A reference cell (like the iodine cell) is critical for 
facility instruments that are not stabilized or for instruments where the user community changes 
the spectrometer configuration. 

During the course of this program, we saw a new state of the art Doppler precision 
emerge from HARPS \citep{M03}, a fiber-fed stable instrument.  We continued our 
efforts to improve precision at Lick; one of us (DAF) wrote a new Doppler code and 
tested new techniques for fitting the data, template deconvolution, and modeling the 
SLSF.  However, even if additional small gains in precision can still be achieved with 
the spectra obtained from the Lick planet search, the bottom line is that high Doppler 
precision (better than 1 \ms) requires that every detail is right, from the instrument 
to the analysis code. The Hamilton spectrograph was not designed to deliver 
sub-meter-per-second precision. To reach much higher precision, the environmental stability 
(pressure and temperature) need to be controlled, the instrument needs to be 
stable against vibration and motion, a fiber scrambler is needed to ensure that the 
illumination of the optics is constant.  It is also likely that significant advances in Doppler precision
will require a new wavelength calibrator that spans most of the observed spectrum 
and does not mask ``stellar noise," the weak signals imprinted in spectral lines from  
signals associated with changes in the stellar magnetic field.

We share this legacy of data and lessons learned 
with the astronomical community and with future planet hunters.

\acknowledgements 
We thank the referee, Artie Hatzes, for helpful comments that have improved this paper. 
We thank the University of California time allocation committee for supporting 
this project for more than twenty years. We acknowledge critical pioneering contributions 
by R. Paul Butler, S. S. Vogt and J. Valenti. We gratefully acknowledge the support and dedication 
of the Lick Observatory staff, in particular, those who worked side-by-side with us for 
two decades: Rem Stone, Tony Misch, Keith Baker, Wayne Earthman, John Morey.  
We thank the many individuals who helped with observing (most
were students at San Francisco State University or the University of California, Berkeley); 
listed roughly in chronological order of their time at Lick, the 
observers included Eric Williams, Chris McCarthy, Phil Shirts, Mario Savio, 
Michael Eiklenborg, David Nidever, Amy Reines, Jason T. Wright, Greg Laughlin, 
Bernie Walp, John Johnson, Eric Nielsen, Melesio Munoz, Howard Isaacson, Julia Kregenow, 
Matt Giguere, Teresa Johnson, Zoe Buck, Katie Peek, John M. Brewer, 
Peter Williams, Raj Sareen, Thomas Ader, 
Kelsey Clubb, Nick van Nispen, Zak Kaplan, Joseph O'Rourke. 
We also thank Thomas Blake at the Environmental Molecular Sciences Laboratory (EMSL) 
for help with new FTS scans of the iodine cell. 
A portion of the research was performed using EMSL, a national scientific user facility 
sponsored by the Department of Energy's Office of Biological and Environmental 
Research and located at Pacific Northwest National Laboratory. 
Fischer acknowledges support by NASA grant NNX10AG08G. Fischer and Spronck thank 
members of The Planetary Society for support of the fiber scrambler project. 
This research has made use of the Simbad database, operated at CDS, Strasbourg, France. 

\facility{Shane} Hamilton spectrograph

\end{document}